\documentstyle[12pt,aps,epsf]{revtex}

\begin{document}
\newcommand{\cross}[1]{#1\!\!\!/}

\newcommand{\vare}{\varepsilon}
\newcommand{\pr}{^{\prime}}
\newcommand{\ppr}{^{\prime\prime}}
\newcommand{\pp}{{p^{\prime}}}
\newcommand{\hp}{\hat{\bfp}}
\newcommand{\hpp}{\hat{\bfpp}}
\newcommand{\hq}{\hat{\bfq}}
\newcommand{\hk}{\hat{\bfk}}
\newcommand{\rx}{{\rm x}}
\newcommand{\rp}{{\rm p}}
\newcommand{\rpp}{{{\rm p}^{\prime}}}
\newcommand{\rk}{{\rm k}}
\newcommand{\rs}{{\rm s}}
\newcommand{\rqq}{{\rm q}}
\newcommand{\bfp}{{\bf p}}
\newcommand{\bfpp}{{\bf p}^{\prime}}
\newcommand{\bfq}{{\bf q}}
\newcommand{\bfx}{{\bf x}}
\newcommand{\bfk}{{\bf k}}
\newcommand{\bfz}{{\bf z}}
\newcommand{\bphi}{{\mbox{\boldmath$\phi$}}}
\newcommand{\balpha}{{\mbox{\boldmath$\alpha$}}}
\newcommand{\bsigma}{{\mbox{\boldmath$\sigma$}}}
\newcommand{\bomega}{{\mbox{\boldmath$\omega$}}}
\newcommand{\bgamma}{{\mbox{\boldmath$\gamma$}}}
\newcommand{\bvare}{{\mbox{\boldmath$\varepsilon$}}}
\newcommand{\intzo}{\int_0^1}
\newcommand{\intinf}{\int^{\infty}_{-\infty}}
\newcommand{\ka}{\kappa_a}
\newcommand{\kb}{\kappa_b}
\newcommand{\lbr}{\langle}
\newcommand{\rbr}{\rangle}
\title{Leading logarithmic contribution to the second-order Lamb shift
induced by the loop-after-loop diagram}
\author{Vladimir A. Yerokhin}
\address{ Department of Physics, St. Petersburg State University,
 Oulianovskaya 1, Petrodvorets, St. Petersburg 198904, Russia\\
and Institute for High Performance Computing and Data Bases, Fontanka 118, St.
Petersburg 198005, Russia}
\maketitle

\begin{abstract}
Contribution of order $\alpha^2 (Z \alpha)^6 \ln^3(Z \alpha)^{-2}$ to the ground-state
Lamb shift in hydrogen induced by the loop-after-loop diagram is evaluated
analytically. An additional contribution of this order is found compared to the
previous calculation by Karshenboim [JETP 76, 541 (1993)]. As a result, an agreement
is achieved for this correction between different numerical and analytical methods.
\end{abstract}

%
%

In the present work a part of the two-loop self-energy correction to the Lamb shift in
hydrogen is investigated, which is induced by the irreducible part of the diagram in
Fig.~1(a). It is referred to as the {\it loop-after-loop} contribution. The
second-order self-energy correction is important for the comparison of the theoretical
prediction with the experimental results for the Lamb shift in hydrogen
\cite{Haensch00} and He$^+$ \cite{Wijng91} and, therefore, it influences the charge
radius of the proton which can be extracted from the Lamb shift in hydrogen
\cite{Karshenboim99}.

The loop-after-loop contribution has been the subject of a recent debate in the
literature. Analytic calculations of its $Z \alpha$-expansion coefficients have been
carried out by Eides and co-workers \cite{Eides93} and Pachucki \cite{Pachucki94} in
order $\alpha^2 (Z \alpha)^5$ and by Karshenboim \cite{Karshenboim93}\footnote{The
same result was re-derived recently in Ref. \cite{Eides00} }  in order $\alpha^2 (Z
\alpha)^6 \ln^3(Z \alpha)^{-2}$. A direct numerical calculation of this correction to
all orders in $Z\alpha$ in the low-$Z$ region was reported by Mallampalli and
Sapirstein \cite{Mallampalli98}. A fit to the data in Ref. \cite{Mallampalli98} is
consistent with the analytical result in order $\alpha^2 (Z \alpha)^5$ but it is in
significant disagreement with Karshenboim's calculation of the leading logarithmic
contribution. As a consequence, a question was raised in Ref.~\cite{Mallampalli98}
about the breakdown of the $Z\alpha$ expansion for the two-loop self energy even for
hydrogen. The subsequent calculation by Goidenko {\it et al.} \cite{Goidenko99}, also
nonperturbative in $Z \alpha$, shows to be compatible with the analytical results.
However, my recent nonperturbative (in $Z\alpha$) calculation \cite{Yerokhin00}
exhibits a good agreement with the results of Ref.~\cite{Mallampalli98} and yields a
corresponding logarithmic contribution which is roughly three times larger than
Karshenboim's result. On the contrary, a recent evaluation by Manohar and Stewart
\cite{Manohar00} supports Karshenboim's value of the leading logarithmic correction
for the total two-loop self energy.

In Refs. \cite{Karshenboim93,Eides00} it is argued that the leading logarithmic
correction for the two-loop self energy is induced only by the diagram shown in Fig.
2(b), if working in the Fried-Yennie gauge. The numerical calculation of
Ref.\cite{Yerokhin00} confirms their value of the corresponding contribution from the
diagram in Fig.~2(b). However, it shows also an additional logarithm cubed with a
coefficient $a_{63} = -0.652(30)$ originating from the diagram presented in Fig.~2(a).
In the present investigation, I derive  the leading logarithmic contribution induced
by this diagram in the case of the ground state of hydrogenlike ions. The resulting
coefficient is found to be $a_{63} = -2/3$, in agreement with my previous numerical
result.


The contribution of the loop-after-loop diagram [Fig.~1(a)] can be written as
\footnote{Relativistic units are used in this Letter ($\hbar=c=m=1$). I use roman
style ($\rp$) for 4-vectors, bold face ($\bfp$) for 3-vectors and italic style ($p$)
for scalars. 4-vectors have the form $\rp = (\rp_0,\bfp)$. The length of the 3-vector
$\bfp$ is denoted by $p$. I also use the notations $\cross{\rp} = {\rm p}_{\mu}
\gamma^{\mu}$, $\hat{\bfp} = \bfp/|\bfp|.$ }
\begin{equation}\label{eq1}
  \Delta E_{\rm lal} =  \frac1{(2\pi)^{12}} \int
  d\bfp d\bfk d\bfk\pr d\bfp\pr
\overline{\psi}_a(\bfp) \Sigma_R(\rp,\rk) G_a^{\rm red}(\rk,\rk\pr)
     \Sigma_R(\rk\pr,\rp\pr) \psi_a(\bfp\pr) \ ,
\end{equation}
where $\psi_a$ is the ground-state wave function, $\Sigma_R$ is the renormalized
self-energy operator, $G_a^{\rm red}$ is the reduced Coulomb Green function, and the
time component of all 4-vectors equals to the ground-state energy $\vare_a$. This
expression is not completely gauge invariant, but it can be shown to be gauge
independent over covariant gauges. In the present investigation the consideration will
be given in the Fried-Yennie gauge. Because of lack of a convenient representation of
the Coulomb Green function, the standard way to evaluate Eq. (\ref{eq1}) is to expand
the inner electron propagators in terms of an interaction with the external nuclear
field. Some of the terms of this expansion are shown in Figs. 2[(a),(b)].

The leading contribution to $\Delta E_{\rm lal}$ is of order $\alpha^2 (Z \alpha)^5$
and  originates from the diagram in Fig.~2(a). It was evaluated in Refs.
\cite{Eides93,Pachucki94} after proving that only relativistic values of the
intermediate momentum are responsible for this correction and that both external
momenta can be put on the mass shell with vanishing space components. The
next-to-leading contribution contains a logarithm cubed. It was derived in Refs.
\cite{Karshenboim93,Eides00} by using an effective-potential approach, where the
self-energy operator is replaced by a potential on the mass shell,
\begin{equation}\label{eq2}
  \Sigma_R(\rp,\rk) \to V_{\rm SE} = -\frac{8\alpha (Z\alpha)}{3} \ln k \ .
\end{equation}
The effective quasilocal potential $V_{\rm SE}$ is defined in a way to generate the
leading contribution to the one-loop Lamb shift. Obviously, this potential leads to a
squared logarithmic contribution in Eq. (\ref{eq1}), after the integration over the
intermediate momentum is carried out within the logarithmic region $Z\alpha \ll k \ll
1$. One can gain one more logarithm if $\ln^2 k$ is integrated with a factor $1/k$,
\begin{equation}\label{eq3}
  \int_{Z\alpha}^1  dk \frac{\ln^2 k}{k} \sim \ln^3 Z\alpha \ .
\end{equation}
In Refs. \cite{Karshenboim93,Eides00}, the factor $1/k$ arises from the second term of
the potential expansion of $G^{\rm red}_a$ in Eq. (\ref{eq1}). This is the reason why
in these considerations Fig.~2(b) yields a logarithm cubed and Fig.~2(a) does not. As
we will see below, the treatment of Refs. \cite{Karshenboim93,Eides00} is incomplete.
A more accurate investigation shows that an analogous factor $1/k$ can be obtained by
taking into account the momentum distribution of the external wave functions in the
diagram in Fig.~2(a).

Let us now derive the next-to-leading contribution to the diagram in Fig.~2(a). While
its leading term can be evaluated assuming the external momenta being put on the mass
shell with vanishing space components, we intend to keep the first-order terms in the
expansion over the space components. The binding energy is still neglected; one can
show that it does not contribute to the order of interest. We write the contribution
of the diagram in Fig.~2(a) as
\begin{equation}\label{eq4}
  \Delta E_{2a} =  \frac1{(2\pi)^9} \int d\bfp d\bfk d\bfpp \psi_a^{\dag}(\bfp)
    V_C(\bfp-\bfk) {\cal K}(\rp,\rk,\rpp) V_C(\bfk-\bfpp)
    \psi_a(\bfpp) \ ,
\end{equation}
where $V_C$ is the Coulomb potential $V_C(\bfq) = -4\pi Z\alpha/q^2$, and $\cal K$ is
the kernel
\begin{equation}\label{eq5}
  {\cal K}(\rp,\rk,\rpp) =\gamma_0 \Lambda^{\rm FY}_{R_0}(\rp,\rk)
    \frac{\cross{\rk}+1}{\rk^2-1} \Lambda^{\rm FY}_{R_0}(\rk,\rpp) \ ,
\end{equation}
with $\Lambda^{\rm FY}_{R_\mu}$ being the renormalized vertex operator in the
Fried-Yennie gauge. For our purposes it is sufficient to keep only the lowest-order
term of the $Z\alpha$ expansion of the Dirac wave function
\begin{equation} \label{eq7}
\psi_{a}(\bfp)
        =\left(  {g_a(p) \chi_{ \ka  m_a}(\hat{\bfp})}
           \atop{ f_a(p) \chi_{-\ka m_a}}(\hat{\bfp}) \right)
    \approx \frac{N}{(p^2+(Z\alpha)^2)^2}\left(  { \chi_{ \ka  m_a}(\hat{\bfp})}
           \atop{ (-p/2) \chi_{-\ka m_a}}(\hat{\bfp}) \right)
           \ ,
\end{equation}
where $N = (2\pi)^{3/2} (32(Z\alpha)^5/\pi)^{1/2}$. In Eq. (\ref{eq7}) and in what
follows, the sign of the approximate equality ($\approx$) is used when terms
irrelevant for the contribution of interest are dropped. Assuming that the external
momentum $\rp$ is close to the mass shell, we can expand the vertex operator over the
space components of $\rp$
\begin{equation}\label{eq6}
  \Lambda_{R_0}^{\rm FY}(\rp,\rk) \approx \Lambda_{0}^{(0)}(\rs,\rk)
    + (\bgamma \cdot \bfp)\Lambda_{0}^{(1)}(\rs,\rk)
    +  (\bfp \cdot \hk) \Lambda_{0}^{(2)}(\rs,\rk)
    + p^2 \Lambda_{0}^{(3)}(\rs,\rk) \ ,
\end{equation}
where $\rs$ is a 4-vector with zero space components, $\rs = (1,{\bf 0})$. The parts
of $\Delta E_{2a}$ induced by the four terms in the righthand side of Eq. (\ref{eq6})
are denoted by $\Delta E_i$, $i = 0\ldots3$.

The correction $\Delta E_0$ can be written by introducing an effective wave function
$\varphi_0$ as
\begin{equation}\label{eq8}
    \Delta E_0 =  \int  \frac{d\bfk}{(2\pi)^3} \varphi_0^{\dag}(\bfk)
        {\cal K}(\rs,\rk,\rs) \varphi_0(\bfk) \ ,
\end{equation}
where
\begin{equation}\label{eq9}
 \varphi_0(\bfk) =  \int \frac{d\bfp}{(2\pi)^3} V_C(\bfk-\bfp) \psi_a(\bfp)
         \approx  -\frac{N}{2k^2} \left( \displaystyle
         {  \chi_{ \ka  m_a}(\hat{\bfk})}
           \atop{\displaystyle
            -\frac{Z\alpha}{2}\, {\rm arctg} \frac{k}{Z\alpha} \,
                    \chi_{-\ka m_a}}(\hat{\bfk}) \right) \ .
\end{equation}
Eq. (\ref{eq8}) contains a contribution of the previous order. To eliminate it, one
should subtract the same expression with the Dirac bispinors replaced by the
Schr\"odinger wave functions. This means that we are interested only in the part of
Eq. (\ref{eq8}) which contains a product of the upper and lower components of
$\varphi_0$. Considering the matrix structure of the kernel ${\cal K}$, one can see
that only the part of ${\cal K}$ proportional to $\gamma_0 (\bgamma \cdot \bfk)$
provides a nonzero contribution to the order of interest. Therefore, we have
\begin{equation}\label{eq10}
  {\cal K}(\rs,\rp,\rs) \approx -\frac{\alpha^2}{(4\pi)^2}
    \frac{\gamma_0 (\bgamma \cdot \bfk)}{k^2}
  {\cal  F}_0(k) \ .
\end{equation}
One can show that ${\cal  F}_0(k) \approx -(112/3) k^2 \ln^2 k^2$ for small $k$.
Keeping in mind that logarithms normally arise from integrals like Eq. (\ref{eq3}), we
can cut off the integration over $k$ in Eq. (\ref{eq8}) by an arbitrary constant
$\Lambda \sim 1$ and use the low-energy asymptotic form for ${\cal  F}(k)$,
\begin{equation}\label{eq11}
  \Delta E_0 \approx -\frac{\alpha^2 (Z\alpha)^6}{2\pi^3} \frac{112}{3}
    \int_0^{\Lambda} dk \frac{\ln^2 k^2}{k} \, {\rm arctg} \frac{k}{Z\alpha}  \ .
\end{equation}
If $Z\alpha$ is set to zero under the integral, it would logarithmically diverge for
small values of $k$. Therefore, $Z\alpha$ can be regarded as an effective cut-off
parameter in this case. We can cut off the $k$-integration by $Z\alpha$ and then set
$Z\alpha$ to zero under the integral
\begin{equation}\label{eq12}
J \equiv \int_0^{\Lambda} dk \frac{\ln^2 k^2}{k} \, {\rm arctg} \frac{k}{Z\alpha}
\approx \frac{\pi}{2}  \int_{Z\alpha}^{\Lambda} dk \frac{\ln^2 k^2}{k}
    \approx \frac{\pi}{12} \ln^3 (Z\alpha)^{-2} \ .
\end{equation}
A rigorous evaluation of the integral $J$ yields the same value of the cubed
logarithmic contribution. So, we have for the correction $\Delta E_0$
\begin{equation}\label{eq13}
  \Delta E_0  \approx \left( -\frac{14}{9} \right) \frac{\alpha^2}{\pi^2}
    (Z\alpha)^6 \ln^3 (Z\alpha)^{-2}   \ .
\end{equation}

The evaluation of the remaining corrections is carried out in a similar way. By a
direct calculation one can show that $\Delta E_1 \approx 0$. The correction $\Delta
E_2$ is written as
\begin{eqnarray}\label{eq14}
  \Delta E_2 &=& -\int \frac{d\bfk}{(2\pi)^3} \frac1{k^2}
                \Bigl\{ \varphi_2^{\dag}(\bfk) \gamma_0 \Lambda_{0}^{(2)}(\rs,\rk)
        (\cross{\rk}+1) \Lambda_{0}^{(0)}(\rk,\rs) \varphi_0(\bfk) \nonumber \\
&&   + \varphi_0^{\dag}(\bfk) \gamma_0 \Lambda_{0}^{(0)}(\rs,\rk)
        (\cross{\rk}+1) \Lambda_{0}^{(2)}(\rk,\rs) \varphi_2(\bfk) \Bigr\} \ ,
\end{eqnarray}
with a wave function $\varphi_2$
\begin{eqnarray}\label{eq15}
   \varphi_2(\bfk) &=&  \int \frac{d\bfp}{(2\pi)^3} (\bfp \cdot \hk)
            V_C(\bfk-\bfp) \psi_a(\bfp) \nonumber \\
         &\approx& -\frac{N}{2k^2} (Z\alpha) \, {\rm arctg} \frac{k}{Z\alpha}
         \left(  { \chi_{ \ka  m_a}(\hat{\bfk})}
           \atop{0}\right) \ .
\end{eqnarray}
Keeping only terms proportional to the product of the upper components, we have
\begin{equation}\label{eq16}
  \Delta E_2  \approx  -\frac{\alpha^2 (Z\alpha)^6}{\pi^3}
    \int_0^{\infty} dk \frac{{\cal F}_2(k)}{k^4} \, {\rm arctg} \frac{k}{Z\alpha} \ ,
\end{equation}
where the function ${\cal F}_2(k)$ can be found to have an asymptotic form ${\cal
F}_2(k) \approx -(64/9)k^3 \ln^2k^2$ for small $k$. Immediately, we have
\begin{equation}\label{eq17}
  \Delta E_2 \approx \frac{64}{9}\, \frac{\alpha^2}{\pi^3} (Z\alpha)^6 J
    \approx \frac{16}{27}\, \frac{\alpha^2}{\pi^2}
    (Z\alpha)^6 \ln^3 (Z\alpha)^{-2}  \ .
\end{equation}
The correction $\Delta E_3$ is evaluated in the same way as $\Delta E_2$ by
introducing a new wave function $\varphi_3$
\begin{eqnarray}\label{eq18}
   \varphi_3(\bfk) &=&  \int \frac{d\bfp}{(2\pi)^3} p^2
            V_C(\bfk-\bfp) \psi_a(\bfp) \nonumber \\
         &\approx& -\frac{N}{k} (Z\alpha) \, {\rm arctg} \frac{k}{Z\alpha}
         \left(  { \chi_{ \ka  m_a}(\hat{\bfk})}
           \atop{0}\right) \ .
\end{eqnarray}
The derivation yields
\begin{equation}\label{der24}
  \Delta E_3 \approx  \frac{8}{27}\,
    \frac{\alpha^2}{\pi^2} (Z\alpha)^6 \ln^3 (Z\alpha)^{-2}  \ .
\end{equation}
Adding all evaluated corrections, we have for the total cubed logarithmic contribution
from the diagram in Fig.~2(a)
\begin{equation}\label{der25}
  \Delta E_{2a} \approx \sum_{i=0}^3 \Delta E_i \approx
  \left( -\frac{2}{3} \right) \frac{\alpha^2}{\pi^2} (Z\alpha)^6 \ln^3 (Z\alpha)^{-2}
     \ .
\end{equation}


The results are commonly represented in the form of a double expansion over
$(Z\alpha)$ and $\ln (Z\alpha)^{-2}$
\begin{equation}\label{d1}
  \Delta E = \left( \frac{\alpha}{\pi}\right)^2 \sum_{ij} a_{ij}\, (Z\alpha)^i \ln^j
  (Z\alpha)^{-2}  \ .
\end{equation}
In this notation the leading logarithmic contribution induced by the diagram in
Fig.~2(a) is found to be $a_{63} = -2/3$, in a good agreement with $a_{63} =
-0.652(30)$ obtained in my previous calculation \cite{Yerokhin00}. This result,
together with the corresponding contribution from the diagram in Fig.~2(b) evaluated
in Ref. \cite{Karshenboim93}, yields the total correction of order $\alpha^2
(Z\alpha)^6 \ln^3 (Z\alpha)^{-2}$ for the loop-after-loop diagram [Fig.~1(a)],
$a_{63}= -2/3-8/27 = -26/27 \approx -0.963$. This value is in good agreement with the
numerical result of Mallampalli and Sapirstein $a_{63} = -0.9$ \cite{Mallampalli98}
and with my numerical calculation $a_{63} = -1.01(8)$ \cite{Yerokhin00}. However, it
strongly disagrees with the evaluation by Goidenko and co-workers \cite{Goidenko99}. A
possible reason for this discrepancy may be a non-covariant renormalization procedure
used in that work, which is known to provide spurious contributions in some cases.
This topic is discussed in more detail in Ref. \cite{Yerokhin00}.

In the present work, only the irreducible part of the diagram in Fig.~1(a) is
investigated. The question if the remaining two diagrams of the two-loop self energy
provide a cubed logarithmic contribution, remains still unclear. An indication that it
could be the case are the results of Pachucki \cite{Pachucki} and Manohar and Stewart
\cite{Manohar00}. The authors calculated the leading logarithmic contribution to the
total two-loop self energy by working in Coulomb gauge \cite{Pachucki} and within the
renormalization-group approach \cite{Manohar00}. Both evaluations agree with
Karshenboim's result. Keeping in mind that we derived the additional contribution from
the diagram Fig.~1(a), one should deduce that the remaining diagrams Figs. 1[(b),(c)]
provide the same contribution with the opposite sign, if working in a covariant gauge.

In summary, I derived an additional cubed logarithmic contribution to the
loop-after-loop correction for the ground state of hydrogenlike ions. As a result, an
agreement between the results of numerical and analytical methods is achieved.
Therefore, there are no grounds at present to question the validity of the $Z\alpha$
expansion for hydrogen. However, the convergence of this expansion for the two-loop
self energy is remarkably slow. As was shown in Ref. \cite{Yerokhin00}, the first
three expansion terms for the loop-after-loop correction cover only about 50\% of the
total result for hydrogen. Thus, it is highly desirable to carry out exact
calculations of the remaining part of the two-loop self energy in the low-$Z$ region.

I would like to thank Prof. Shabaev for his guidance and continued interest during the
course of this work. Valuable conversations with S. G. Karshenboim are gratefully
acknowledged. I thank also M. I. Eides, L. N. Labzowsky, and Th. Beier for interesting
discussions. This work was supported in part by the Russian Foundation for Basic
Research (Grant No.~98-02-18350) and by the program "Russian Universities. Basic
Research" (project No. 3930).


\newpage
\begin{figure}
\centerline{ \mbox{ \epsfxsize=\textwidth \epsffile{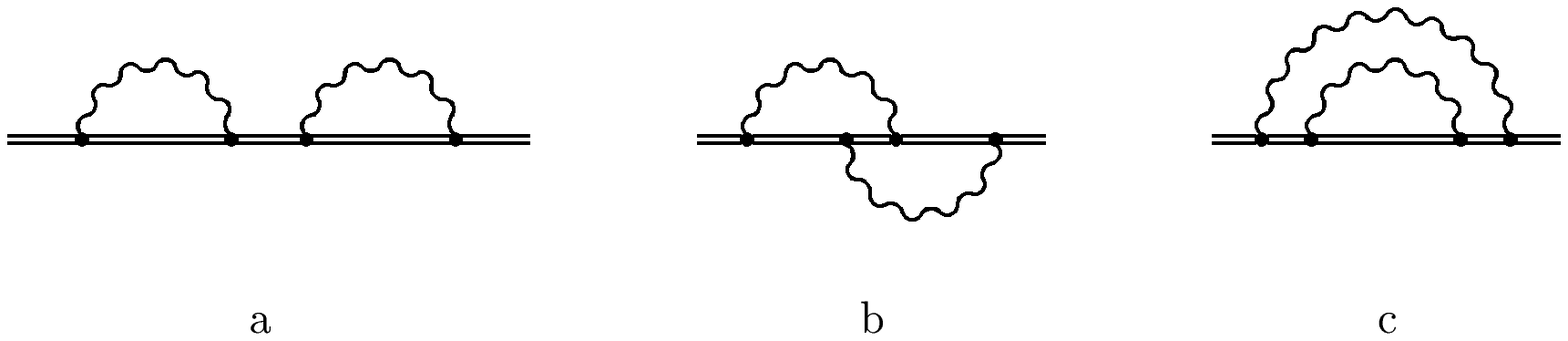} }} \caption{One-electron
self-energy Feynman diagrams of second order in $\alpha$.} \label{se2o}
\end{figure}

\begin{figure}
\centerline{ \mbox{ \epsfxsize=\textwidth \epsffile{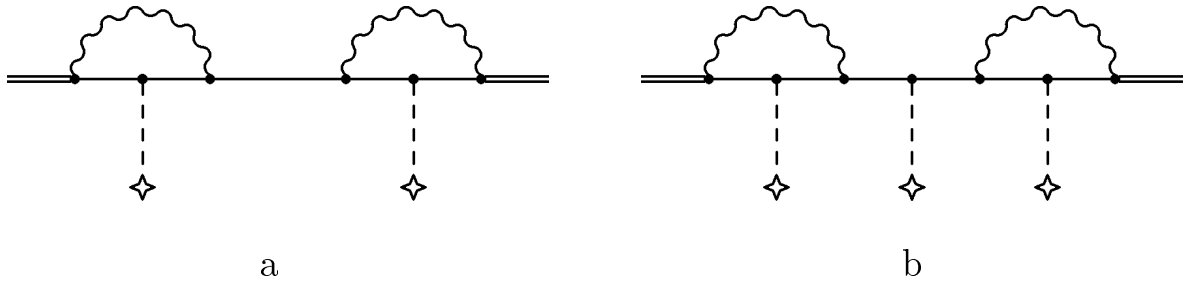} }} \caption{Two terms
of the potential expansion of the diagram in Fig.~1(a). A double line denotes an
electron in the field of the nucleus. A single line indicates a free electron. A
dashed line denotes a Coulomb interaction with the nucleus.}
\end{figure}

\end{document}